# Paramagnetic alignment of fractal grains

A. Lazarian

DAMTP, Silver Street, University of Cambridge, UK.
E-mail: al126@uk.ac.cam.amtp

**Abstract.** Paramagnetic alignment of fractal suprathermally rotating grains is discussed. It is shown that if the concentration of $H_2$ formation sites is low and resurfacing is active, fractal structure of grains enhances their alignment. Studying the influence of grain surface physics and chemistry on the alignment we found that there exist two critical values of grain sizes, and the alignment of smaller grains is expected to decrease as compared to the predictions of the Purcell's theory (1979). One of the critical sizes is due to intensified poisoning of active sites, while the other is a result of a finite value of the imaginary part of magnetic susceptibility in the limit of high angular velocities. On the contrary, if active sites completely cover grain surface, suprathermal rotation, and therefore efficient alignment, is possible only for a limited range of grain sizes.

## 1. Introduction

Although the starlight polarization by the ISM dust was first reported as far back as late 40s (see Hiltner 1949) it is still a theoretical challenge to explain the alignment of dust grain that causes it. The most widely accepted explanation of this phenomenon as being due to paramagnetic relaxation (Davis & Greenstein 1951) was later criticised (Jones & Spitzer 1967) as inadequate to account for observations unless enhanced imaginary part of grain magnetic susceptibility is assumed (see also Duley 1978, Sorrell 1994). This assumption became a corner stone of the Mathis (1986) theory of alignment of superparamagnetic grains. Another attempt to overcome the said drawback was made in Purcell (1975, 1979), where spinning of grains up to suprathermal velocities was considered to diminish the influence of thermal disorientation. These two modifications of the paramagnetic mechanism are widely believed to be the strongest candidates for explaining starlight polarization, though there are indications that Gold-type processes (Gold 1951, 1952, Dolginov & Mytrophanov 1976, Lazarian 1994a) are rather common in the ISM.

In this paper we concentrate our attention on the Purcell's (1979) modification of the paramagnetic mechanism. For grains approximated by regular ellipsoids, it was shown (Spitzer & McGlynn 1979) to improve the alignment only marginally if high rates of resurfacing were assumed. Here we study physics and chemistry of resurfacing and the changes that a fractal structure of grains entails. The assumption of porous irregular grain structure, which can be described using a notion of fractality, corresponds to the present-day state of our knowledge (see Wright 1987, Whittet 1992, Williams 1993, Lume & Ranhola 1994).

The structure of this paper is as follows. The concept of fractality in application to grains is briefly discussed in section 2, while the suprathermal rotation of grains due to recoils from $H_2$ molecules is studied in section 3. Then in section 4, we address the paramagnetic mechanism of grain alignment and the effect of grain fractality on the measure of alignment. In section 5, theoretical predictions are analysed in the light of observational data available.

## 2. Fractal grains

Whatever the cause of grain formation they are expected to be irregular. So are the species found in the stratosphere (see fig. 7.3 in Williams 1993) and those formed in an arc (Codina-Landaberry et al. 1974). In fact, it is believed that ISM grains may consist of small particles jumbled together loosely (Mathis & Whiffen 1989). Such grains 'are the natural result of coagulation and disruption of grains as they cycle into clouds' (Mathis 1990).

Therefore it seems natural to assume that ISM grains are fractal (Mandelbrot 1983). This idea was advocated in Wright (1987) on the basis of the long-wavelength adsorption data, and here we are going to study its implications for the alignment.

To remind our reader a few facts about fractals, we use probably the most familiar example of fractals related to the length of a coastline. This length $G$ depends on the scale of measurements $G_0$ as

$$G \sim G_0^{T-d} \qquad (1)$$

where $d$ is the fractal dimension, and $T$ is the topological dimension, which is 1 for the coastline. Turning from the coastline to grains, we consider the surface of a well defined volume $V_0$. In this case, one may state that two grains are similar if for them the ratio $S_f^{1/d}(r_0)/V_0^{1/3}$ is the same, where $S_f(r_0)$ is the grain surface area measured with the yardstick $r_0$. If grains consist of small spheres, the size of the latter determines $r_0$. The analogy between a coastline and grains can be protracted as although in both cases we do not know all the details of their formation we are still able to define their dimension.

Further on for the sake of convenience, we will use $l$ for the dimension of grains (a diameter for spherical species) instead of $V_0^{1/3}$. An important property of fractal grains that influences our results is that $S_f \sim l^d$. The actual value of $d$ may be calculated within an adopted model. If the spheres of size $r_0$ are attached randomly one by one to a growing cluster, their number grows as $l^{2.5}$ (Witten & Cates 1986). Assuming that the area of intersection is small, it is easy to see that

$$S_f = \pi l^2 \left(\frac{l}{r_0}\right)^{d-2} \qquad (2)$$

where $d = 2.5$ should be put for the latter case. However, if the spheres themselves are porous, the actual surface can be greater. The fractal dimension in its turn can change with the scale that we study. If grains are being formed from loosely connected PAH structures (Duley & Williams 1993), the ratio $l/r_0$ may reach $10^3$ for grains of $l \sim 3 \cdot 10^{-5}$ cm. Yet, we will use more conservative estimates in our paper.

In short, grains being three dimensional objects can have surface growing faster than $l^2$. In contrast, if grains are formed by chains of spheres sticking together, these loosely complexes can have dimension less than 2 (Wright 1987). Later we will be mostly concerned with the three dimensional clusters, as we believe that very loose complexes are not strong enough to withstand suprathermal rotation.

## 3. Suprathermal rotation of fractal grains

The concept of suprathermal rotation was first introduced in Purcell (1979). Various processes, e.g. variations of the accommodation and photo-emission coefficients, variations of the number of sites of $H_2$ formation and the difference in cross-sections for left- and right-hand circularly polarized quanta can result in suprathermal rotation of grains. An important point (see Lazarian 1994b) is that the processes associated with corpuscular and radiative fluxes bring about the Gold-type alignment which, as a rule, has a much shorter time-scale. Therefore for the rest of the paper we will consider rotation induced by the $H_2$ formation over the grain surface. These reactions are expected to proceed with a rather high efficiency: almost every $H$ atom adsorbed by the surface must leave it as part of $H_2$ molecule (see Williams 1988). The surface concentration of active sites varies for different grain species.

We begin with considering low density distribution of active sites. As a rule, the sites are subdivided into two categories: sites of physical adsorption with the energy less than 0.5 eV and those of chemical adsorption with the energy of $\sim 0.5 - 5$ eV (Watson 1976). Although only a few sites of chemical binding per grain are required to account for the observed quantities of $H_2$ (Hollenbach & Salpeter 1970, 1971), some authors believe that their number is much greater and corresponds to the density of $\alpha_{H2} = 10^{15}$ cm$^{-2}$ (see Tielens & Allamandola 1987).

Due to quantum tunnelling, adsorbed $H$ atoms acquire high mobility (Watson 1976) and form $H_2$ molecules both over external grain surface and within numerous pores of a fractal grain. The effects of these two types of reactions upon the grain are different. It is obvious that if an active site is located well within a narrow pore, the ejected $H_2$ molecule has much chances to thermolise by colliding with the pore walls before leaving the grain. The entire energy of the reaction is expected to be transferred to the grain in the form of heat and the recoil from such $H_2$ ejection is negligible. For active sites on the external grain surface, a recoil from every departing molecule is $m_{H2}v_{H2} = 2(m_{H2}E)^{1/2}$, where $E$ is the kinetic energy of the ejected $H_2$ molecule $\sim 0.2$ eV (Williams 1988). The latter value intimately depends on the chemical composition of grain surface, and this can bring about differences in grain alignment for different species (Lazarian 1994c). The situation is intermediate for grains within pores but close to the external surface. They are expected to experience a few collisions with pore walls before leaving the grain and the residual energy of the molecule depends on the details of its interaction with the grain surface in the course of collisions, e.g. on the accommodation coefficient. This includes first of all the portion of chemical energy converted into $H_2$ molecule translational motion. Apart from this energy channel, there are other possibilities, e.g. the transfer of vibrational energy of a newly formed $H_2$ molecule to vibrational modes of the grain. Because of the inherent quantum nature of this process the contributions to different channels depend on the spacing of the energy levels involved as well as on the nature of the interaction. The estimates (see Lucas & Ewing 1981) indicate that a relatively small part of recombination energy is transferred into translational motion. The ejected $H_2$ molecules can still be vibrationally 'hot', and this may influence their later interactions with the grain. Indeed, a molecule colliding with a grain loses a fraction of its translational energy that depends on the ratio of the Debye frequency $\omega_D$ and the inverse time-scale of the collision $\omega_0$ (Watson 1976)

$$\omega_0 = \left(\frac{E+E_b}{2mb^2}\right)^{1/2} \qquad (3)$$

where $E_b$ is the binding energy of $H_2$ molecule, and $b$ is the length scale for the repulsive interaction $\sim 10^{-7}$ cm. For $\omega_0 \ll \omega_D$, the energy transfer to the grain is $\sim (E+E_b)\omega_0^3\omega_D^{-3}$ (Hollenbach & Salpeter 1970), which enables a molecule to collide $\sim \omega_0^3\omega_D^{-3}$ times with the grain before losing its translational energy. At the same time, one cannot exclude the possibility of converting some additional internal energy of 'hot' $H_2$ molecules into translational energy during these interactions (Watson 1976). This can make internal active sites well within the body of a fractal grain efficient for ejecting high speed $H_2$ molecules through its pores.

For simplicity, we consider that molecules formed within pores up to $l_{ac}$ depth from the grain outer boundary retain their energy. For fractal porous grains the core of scale $l - l_{ac}$ is passive in terms of contributing to the recoils from $H_2$ molecules, although $H_2$ formation is expected to proceed within this 'passive' core nearly with the same rate as over the outer surface. Among other factors, $l_{ac}$ depends on the pore geometry. For pores approximated by cylinders, it is only the normal component of velocity that decrease due to molecular collisions with the pore walls. Integrating over a Gaussian distribution of emerging molecules with the mean energy $E$, it is possible to show that the grain momentum gained as a result of emission of $n_o$ $H_2$ molecules through such a pore is equal to $n_o (Em_{H2}/3\pi)^{1/2}$ and the torque is equal to $n_oE/3$. Note, that we considered a pore with one end closed and only the velocity component along the pores axis preserved. The torque is less if the pore is irregularly twisted.

It was shown in Purcell (1979) that internal dissipation of energy within grains, mainly due to the Barnett relaxation,[1] suppresses rotation around any axis but the axis of the greatest inertia on the time-scale $\sim 10^7/\eta$ s, where $\eta$ is the ratio of grain rotational energy to the equipartition energy $\sim kT$. Thus grains rotate around their major axes of inertia that we denote here $z$-axis. Therefore the problem becomes one dimensional as the two other components of the torque contribute only to insignificant nutations. The number of $H_2$ molecules ejected per second from an individual site is $\sim \gamma_1 l^2 v_1 n_H \nu^{-1}$, where $\gamma_1$ is the portion of $H$ atoms with velocity $v_1$ and concentration $n_H$ adsorbed by the grain, while $\nu^{-1}$ is the number of active sites over the surface grain. Then the mean square of a residual torque is

$$\langle [M_z]^2 \rangle \approx \frac{\gamma_1^2}{32} l^6 n_H^2 m_{H2}^2 v_1^2 E \nu^{-1} \left(\frac{l_{ac}}{l}\right)^{2d} \qquad (4)$$

which spins up grains to angular velocities limited only by friction forces, i.e.

$$\Omega = \langle [M_z]^2 \rangle^{1/2} \frac{t_d}{I_z} \qquad (5)$$

where $I_z$ is the $z$ component of the momentum of inertia, and $t_d$ is the rotational damping time (Spitzer & McGlynn 1979):

$$t_d \approx 0.6 t_m = 0.6 \frac{l \varrho_s}{n m v_1} \qquad (6)$$

where $t_m$ is the time that takes a gain of density $\varrho_s$ to collide with gaseous atoms of the net mass equal to that of the grain, $n$ and $v_1$ denote the density and velocity of atoms, respectively, while $m$ is the mass of an individual atom. Therefore

$$\Omega \approx \frac{\gamma_1}{4\nu^{1/2}} \left(\frac{n_H}{n}\right) \left(\frac{m_{H2}}{m}\right) \left(\frac{l_{ac}}{l}\right)^d \frac{v_{H2}}{l} \qquad (7)$$

which is

$$\Omega \approx 1.4 \cdot 10^8 \left(\frac{\gamma_1}{0.2}\right) \left(\frac{n_H}{n}\right) \left(\frac{m_{H2}/m}{2}\right) \left(\frac{2 \cdot 10^9 \text{ cm}^{-2}}{\alpha_{H2}}\right)^{\frac{1}{2}}$$
$$\times \left(\frac{v_{H2}}{4 \cdot 10^5 \text{ cm s}^{-1}}\right) \left(\frac{2 \cdot 10^{-5} \text{ cm}}{l}\right)^{2+(d-2)/2}$$
$$\times \left(\frac{r_0}{10^{-6} \text{ cm}}\right)^{\frac{d-2}{2}} \left(\frac{l_{ac}}{l}\right)^d \text{ s}^{-1} \qquad (8)$$

if we assume that a grain of $l = 2 \cdot 10^{-5}$ cm has $\nu = 100$ active sites and $d = 2.5$. If active sites completely cover the grain $\alpha_{H2}$ becomes $\sim 2 \cdot 10^{15}$ (see Tielens & Allamandola 1987), and $\Omega$ reduces to $1.4 \cdot 10^5$ s$^{-1}$ which is comparable with the thermal rotational velocity of grains for $T_g = 90$ K and $T_s = 10$ K

$$\omega_T \approx 9 \cdot 10^4 \left(\frac{2 \cdot 10^{-5} \text{ cm}}{l}\right)^{\frac{5}{2}} \left(\frac{0.5(T_g + T_s)}{50 \text{ K}}\right)^{\frac{1}{2}} \text{ s}^{-1} \qquad (9)$$

and for grains with $l \approx 2 \cdot 10^{-6}$ cm, $\omega_T$ becomes greater than $\Omega$. If $l_{ac} < l$, thermal rotation dominates for a wide range of grain sizes.[2]

---

[1] A quantitative treatment through solving Fokker-Plank equations is given in Lazarian (1994a).
[2] The mean temperature in Eq. (9) is used because atoms thermolise over grain surface after non-elastic collisions.

Note, that fractal grains with small $l_{ac}/l$ ratio are likely to lose their suprathermal advantage. This corresponds to the situation when the absolute majority of $H_2$ molecules thermolise before leaving the grain. The $H_2$ thermalization depends on the material of the grain, its size and the structure of its fractal maze.

The implications of the increase of the number of active sites may be dramatic. It is possible to show that for a regularly shaped fractal grain with a uniform distribution of active sites chemically driven rotation corresponds to temperature

$$T \approx T_a \left\{ 1 + \frac{\gamma_1}{2} \frac{E}{kT_a} \left(\frac{l_{ac}}{l}\right)^d \right\} \qquad (10)$$

where $T_a$ is the gas temperature, and $k$ is the Boltzman constant. For $T_a = 80$ K and $l_{ac}/l = 0.3$, this gives just 100 K, which in fact nullifies the effect of suprathermality. To preserve the said effect, we may allow for the change of the density of *effective* active sites per unit of grain geometric surface $\alpha_{H2}$ over patches of area $\epsilon_{H2}$.[3] The corresponding variation

$$\vartheta_{H2}^2 = \left\langle \left(\frac{\alpha_{H2}}{\langle \alpha_{H2} \rangle} - 1\right)^2 \right\rangle \qquad (11)$$

will be used further. We emphasise *effective*, since only active sites in pores within the depth $l_{ac}$ are considered. As the number of pores over grain surface varies, this gives a natural explanation for variations of $\alpha_{H2}$.

If we ignore the impacts from $H$ atoms, the pressure due to the recoils from $H_2$ molecules is

$$P = \frac{\gamma_1 \alpha_{H2} m_H n_H v_{H2} v_1}{\langle \alpha_{H2} \rangle} \left(\frac{l_{ac}}{l}\right)^d \qquad (12)$$

where $m_H n_H v_1$ is the mass flux of $H$ atoms. Therefore the torque applied to the grain due to the variation of $\alpha_{H2}$ over a patch $\epsilon_{H2}$ is proportional to

$$\frac{1}{\langle \alpha_{H2} \rangle} \epsilon_{H2}^2 \frac{\gamma_1}{2} (\alpha_{H2} - \langle \alpha_{H2} \rangle) m_H n_H v_{H2} v_1 l \left(\frac{l_{ac}}{l}\right)^d \qquad (13)$$

and the mean square of the torque is

$$\langle [M_z]^2 \rangle \approx \frac{\gamma_1^2}{4} m_H \epsilon_{H2}^2 n_H^2 v_1^2 E \vartheta_{H2}^2 l^4 \left(\frac{l_{ac}}{l}\right)^{2d} \qquad (14)$$

where it is taken into account that the number of $\epsilon_{H2}$ patches is $\left(\frac{l}{\epsilon_{H2}}\right)^2$. The mean angular velocity of grains is

$$\Omega \approx 0.3 \gamma_1 \left(\frac{l_{ac}}{l}\right)^d \left(\frac{\epsilon_{H2} \vartheta_{H2}}{l}\right) \left(\frac{n_H}{n}\right) \left(\frac{m_{H2}}{m_a}\right) \left(\frac{v_{H2}}{l}\right) \qquad (15)$$

which gives $\sim 5 \cdot 10^5$ s$^{-1}$ for $n_H = n$, $l_{ac} = l$ and $\left(\frac{\epsilon_{H2} \vartheta_{H2}}{l}\right) = \frac{1}{300}$.

Assuming that $\vartheta_{H2}$ is proportional to $l^{1/2}$, it is easy to see that $\Omega \sim l^{-3/2}$ and therefore at sizes $\sim 5 \cdot 10^{-6}$ cm the thermal effects become dominant. Similarly, for large $l$, a greater portion of molecules thermolise before leaving deep pores of grains and rotation cease to be suprathermal.

---

[3] Naturally, a more complete description of surface variations may be given by employing a distribution of scale lengths, i.e. a spectrum of spatial frequencies. In our model, we avoid these complications.

Therefore if fractal grains are completely covered by active sites, they can be suprathermal due to $H_2$ formation within a limited range of sizes $[l_1, l_2]$; for grains larger than $l_2$ and smaller than $l_1$ thermal effects dominate. Whether or not these ranges overlap as well as the precise values of $l_1$ and $l_2$ depend on the grain composition and structure. We intend to address this problem in our future research. Note that it is due to grain fractality that the upper limit $l_2$ appears; for solid smooth grains covered with active sites only the lower limit $l_1 \sim 10^{-6}$ cm is expected.

## 4. Paramagnetic mechanism

### 4.1. Factors of alignment

A paramagnetic grain rotating in the external magnetic field **B** is subjected to dissipative torques (Davis & Greenstein 1951, see also Jones & Spitzer 1967, Purcell & Spitzer 1971, Roberge et al. 1993). In the grain coordinate system, the component of **B** perpendicular to the $z$ axis appears as a rotating field. If the angle between **B** and **z** is $\theta$, the magnitude of the said component is $B \sin \theta$. The magnetic moment **M** produced by the imaginary part of the magnetic susceptibility $\chi''$ is perpendicular to both **z** and **B** and has the magnitude $V \chi'' B \sin \theta$. The retarding torque $\mathbf{M} \times \mathbf{B}$ causes rotation of the vector of angular momentum $\mathbf{J} = I\mathbf{\Omega}$ towards the direction of **B**, namely,

$$\frac{d\theta}{dt} = \frac{V \chi'' B^2 \sin \theta}{I_z \Omega} \quad (16)$$

where it is taken into account that the rotation is around the axis of maximal inertia (i.e. $I = I_z$).

For low frequencies, $\chi''$ is given by (Spitzer 1978)

$$\chi'' \approx 2.5 \times 10^{-12} \frac{\Omega \cos \theta}{T_S} \quad (17)$$

which holds up to frequencies $\omega_c = 4 \times 10^{11} N_e \mu^2 / k$, where $\mu = 0.9 \times 10^{20}$ erg G$^{-1}$ is the electron magnetic moment and $N_e$ is the number of unpaired electron spins per cm$^3$. For a grains with one paramagnetic ion per hundred atoms, $\omega_c \approx 10^7$ s$^{-1}$. If $\Omega \gg \omega_c$, which may be a case for suprathermally rotating grains, $\chi''$ becomes independent of $\Omega$ and approaches the static susceptibility. Therefore at frequencies $10^8 - 10^9$ s$^{-1}$, $\chi''$ is expected to come to saturation and not to depend on $\Omega$. Note, that the dependence $T_S^{-1}$ is preserved for these saturation values.

In short, large angular velocities of grains increase the characteristic time of paramagnetic relaxation as $\chi''$ comes to its asymptotic values (compare Whittet 1992). The characteristic time of this relaxation is

$$t_r \approx \frac{10^{14}}{q} \left( \frac{l}{2 \cdot 10^5 \text{ cm}} \right)^2 \left( \frac{T_S}{15 K} \right) \left( \frac{2 \cdot 10^{-6} \text{ G}}{B} \right)^2$$
$$\times K(\Omega) \left( \frac{\varrho}{3 \text{ g cm}^{-3}} \right) \text{ s} \quad (18)$$

where $T_S$ is the temperature of grain material, $\varrho$ is its density, $q$ is the factor of imaginary part of magnetic susceptibility enhancement (Jones & Spitzer 1967), and $K(\Omega)$ is a function with the following asymptotics

$$K(\Omega) = \begin{cases} 1, & \Omega \ll \omega_c \\ \frac{\Omega}{\omega_c}, & \Omega \gg \omega_c \end{cases} \quad (19)$$

For our model, we will consider that transition from one asymptotic to another is at $\Omega = \omega_c$.

For the original Davis-Greenstein process, $t_r$ should be compared with the time that takes a grain of mass $m$ to collide with gas atoms of the same total mass

$$t_m \approx 2 \cdot 10^{12} \left( \frac{l}{2 \cdot 10^5 \text{ cm}} \right) \left( \frac{\varrho}{3 \text{ g cm}^{-3}} \right) \left( \frac{1.3 \cdot 10^5 \text{ cm s}^{-1}}{v_a} \right)$$
$$\times \left( \frac{20 \text{ cm}^{-3}}{n} \right) \left( \frac{10^{-24} \text{ g}}{m_a} \right) \quad (20)$$

where the velocity $v_a$ corresponds to that of $H$ atoms at $T \approx 80$ K. Note, that in many cases the velocity $v_a$ is dominated by Alfvénic phenomena (see Aron & Max 1975, Elmegreen 1992, Myers 1987, 1994, Lazarian 1992, 1993, 1994a), the fact that makes $t_m$ even smaller. The ratio of $t_m$ to $t_r$

$$\delta_1 \approx 0.02 q \left( \frac{2 \cdot 10^{-5} \text{ cm}}{l} \right) \left( \frac{15 \text{ K}}{T_S} \right) \left( \frac{B}{2 \cdot 10^{-6} \text{ G}} \right)^2$$
$$\times \left( \frac{1.3 \cdot 10^5 \text{ cm} \cdot \text{s}^{-1}}{v_a} \right) \left( \frac{20 \text{ cm}^{-3}}{n} \right) K^{-1}(\Omega) \quad (21)$$

determines the mean value of precession angle $\beta$ and therefore the measure of grain alignment is

$$\sigma = \frac{3}{2} \left\langle \cos^2 \beta - \frac{1}{3} \right\rangle \quad (22)$$

According to J. Mathis (private communication), the low value of $B$ in Eq. (21) should be treated as standard and 'there might be as much as a fractor 10 or more increase in $\delta_1$ for the $B^2$ term (see Boulares & Cox 1990, fig. 3b)'. It is possible to show that for the Davis-Greenstein process $\sigma$ takes the following form (Aannestad & Greenberg 1983)

$$\sigma = \frac{2}{3} \frac{1}{1-\xi^2} \left[ 1 - \frac{\xi}{\sqrt{1-\xi^2}} \arcsin \sqrt{1-\xi^2} \right] - \frac{1}{2} \quad (23)$$

where

$$\xi^2 = \frac{1 + \delta_1 \frac{T_S}{T_a}}{1 + \delta_1} \quad (24)$$

The corresponding plot in Fig. 1 indicates that it is only through an appreciable increase of $q$, as compared to the usual materials, (see Jones & Spitzer 1967) that one can reconcile the mechanism with observations. Although not impossible (see Sorrell 1994), the above conjecture needs additional assumptions (Mathis 1986) to explain the polarization-wavelength dependence (see Wilking et al. 1982). There are also indications that the modified theory does not fit the UV observations (see Clayton et al. 1992).

Thus, the suggestion that a better alignment can be delivered by suprathermally rotating grains (Purcell 1975, 1979) was met enthusiastically in the community (see Martin 1978, Johnson 1982). Indeed, if grain rotational energy is greater than $k$ times any temperature in the system, chaotic bombardment of ambient gas cannot alter the regular rotation. Thus, if the time associated with a regular torque (see section 3) is $t_L$, it is this time that should be compared with $t_r$. In general, $t_L$ may be much greater than $t_m$ and therefore it gives a chance of improving alignment without forcing $t_r$ down. It was argued in Purcell (1979) that the quasi-empirical formula

$$t_x = 1.3(t_L + 0.6 t_m) \qquad (25)$$

is a good approximation within the accuracy of 20% for the 'mean time back to crossover' and this should be compared with $t_r$. Indeed, if $t_L \ll t_m$, the applied torques are effectively chaotic and $t_x$ should be $\sim t_m$. In the opposite case ($t_L \gg t_m$), which is much more interesting in the view of the above discussion, $t_x$ should be of the order of $t_L$ which is also satisfied. Thus, although Eq. (25) is not a precise theoretical result and was only tested for $0.1 < t_L/t_m < 10$ (Purcell 1979) we will use it for the whole range of $t_L$ that we encounter further in this paper.

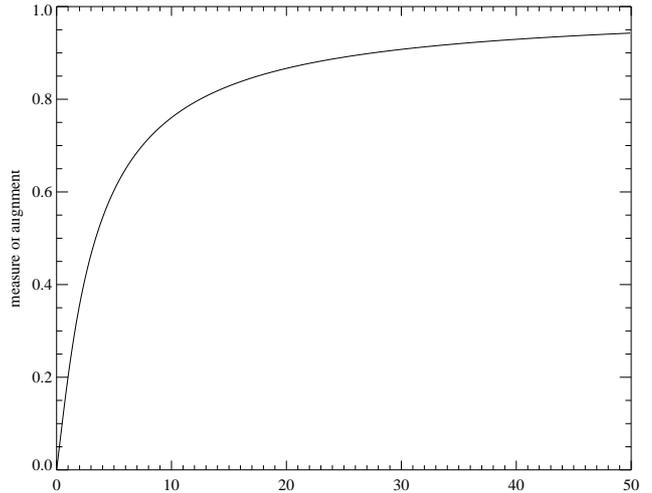

**Fig. 1.** The measure of the Davis-Greenstein alignment as a function of grains size measured in $10^{-5}$ cm units for different factors $q$.

In was shown in Spitzer & McGlynn (1979) that if the disorientation is not complete during a crossover the following parameter $\delta_{\mathit{eff}}$ enters the theory:

$$\delta_{\mathit{eff}} = \begin{cases} \frac{t_x}{t_r}, & F > 1 \\ \frac{t_x}{F t_r}, & F < 1 \end{cases} \qquad (26)$$

where $F$ is the 'disorientation parameter', which is discussed further. The mean alignment measure [see Eq. (22)] may be expressed using $\delta_{\mathit{eff}}$ (Aannestad & Greenberg 1983)

$$\langle \sigma \rangle = \delta_{\mathit{eff}}^{-1} \int_0^\infty \sigma(\delta) \exp\left\{-\frac{\delta}{\delta_{\mathit{eff}}}\right\} d\delta \qquad (27)$$

where $\sigma(\delta)$ is given by (Purcell 1979)

$$\sigma(\delta) = \frac{3}{2} \frac{1 - (\exp \delta - 1)^{-1/2} \arctan(\exp \delta - 1)^{1/2}}{[1 - \exp(-\delta)]} - \frac{1}{2} \qquad (28)$$

This measure is shown in Fig. 2 as a function of $\delta_{\mathit{eff}}$.

### 4.2. Spin-up for grains

It follows from the above that to account for the alignment when $t_r$ is given by Eq. (18), one needs to have $t_L \gg t_m$. The difficulty on this way is the finite life time of active sites over grain surface.

**Fig. 2.** The measure of suprathermal alignment as a function of $\delta_{\mathit{eff}}$.

Active sites of $H_2$ formation are believed to be associated with places of chemical binding. $H$ atoms physically adsorbed by a grain (the binding energy $< 0.1$ eV) migrate over its surface until they meet either a vacant site for chemisorption or a trapped chemisorbed $H$ atom and form a molecule (see Tielens & Allamandola 1987).

A process that can limit the life of active sites is resurfacing due to accretion of a new monomolecular layer (Spitzer & McGlynn 1979). The accretion rate is proportional to the incoming atomic flux times overall grain cross-section $S(l) \sim \pi l^2$. Therefore a grain of mass $m(l)$ grows as

$$\frac{dm(l)}{dt} = \frac{1}{4} S(l) v_a n_a m_a \xi_a \qquad (29)$$

where $\xi_a$ is the adsorption probability, a coefficient with the range of values $0 < \xi_a < 1$ for atoms of density $n_a$, velocity $v_a$ and mass $m_a$. Assuming that accreting atoms uniformly cover the entire physical surface of grains and substituting $dm(l) = \varrho S_f(l) dl$, where $S_f(l)$ is given by Eq. (2), one obtains for the accreting time:

$$t_a = 4 t_m \left(\frac{\triangle h}{l}\right) \left(\frac{n_H}{n_a}\right) \left(\frac{m_H}{m_a}\right) \left(\frac{v_H}{v_a}\right) \left(\frac{l}{r_0}\right)^{d-2} \frac{1}{\xi_a} \qquad (30)$$

where $\triangle h$ is the thickness of an adsorbed layer. If $v_H$ and $v_a$ are due to thermal motions, then $\left(\frac{v_H}{v_a}\right) = \left(\frac{m_a}{m_H}\right)^{1/2}$ and $t_a$ becomes proportional to $\left(\frac{m_H}{m_a}\right)^{1/2}$. However, if non-thermal Alfvénic motions dominate, $v_H$ becomes equal to $v_a$ and $t_a$ is less or proportional to $\left(\frac{m_H}{m_a}\right)$. The ratio $(l/r_0)^{d-2}$, where $r_0$ is the size of an elementary sphere our model grain is composed of, appears due to assumed high mobility of accreted species. Whereas if poisoning atoms hydrogenate on their arrival, the products like $H_2O$ are essentially immobile over grain surface. In such a case, accretion is active only over external grain surface. However, the life time of active sites within grain pores is expected to increase as compared with that discussed in Spitzer & McGlynn (1979).

We believe that oxygen is the most important agent poisoning the active sites. Other chemical elements either form compositions that desorb more easily than $H_2O$ or have a negligible abundance in the gaseous phase. As all heavy elements oxygen is usually depleted in the gaseous phase. For instance, its depletion towards $\zeta$ Persei is -0.5, which gives $n_O/n_H \approx 10^{-3.5}$ (Cardelli et al. 1991). Assuming that a layer of $\triangle h \approx 3.7 \cdot 10^{-8}$ cm effectively covers active sites over grain surface (Aannestad & Greenberg 1983), one obtains

$$\left(\frac{\triangle h}{l}\right) = 1.85 \cdot 10^{-3} \left(\frac{0.2 \cdot 10^{-5} \text{ cm}}{l}\right) \tag{31}$$

Therefore for thermal accretion one has

$$t_a \approx 4 t_m \left(\frac{0.2 \cdot 10^{-5} \text{ cm}}{l}\right) \left(\frac{l}{r_0}\right)^{d-2} \frac{1}{\xi_a} \tag{32}$$

considering the accretion of oxygen atoms. The estimate for $t_a$ can be further increased if one assumes that $\xi_a < 1$ and the desorption is present. The value of $\xi_a \sim 0.1$ was used in Spitzer (1978, p. 208) for illustrative purposes while values 0.02 and 0.2 were reported for sticking probabilities of $C$ over silicate and graphite covered with a monolayer of $H_2O$ surfaces (see Williams 1993). It was shown in Jones & Williams (1984) that the effective sticking probability varies as mantles begin to be deposited and as a consequence the time of establishing monolayer coverage may be increased by a factor of about three.

The products of grain surface chemistry are also subjected to desorption which increases the time of resurfacing. According to Jones & Williams (1984) the probability of retaining the products $OH$ and $H_2O$ on the grains surface is not less than 0.7 in dark quiescent clouds, whereas $NH$ may be predominantly released on formation (see Wagenblast et al. 1993). Immediately after formation the reaction products are subjected to photodesorption. This process can be driven not only by UV quanta (see Johnson 1982), but also by water ice 3 $\mu$m absorption (Williams et al. 1992). Note, that 3 $\mu$m radiation penetrates much further into dark clouds. The energy release associated with $H_2$ formation may be another cause of desorbing weakly bound molecules such as $CO$ in the vicinity of active sites (Duley & Williams 1993). This means that active sites are able to control to some extent resurfacing in their vicinity.

In brief, we may expect to obtain $t_a$ of the order of $10 \; t_m (l/r_0)^{d-2}$ using rather conservative estimates for the above parameters.

The assumption of a uniform coverage of grain surface by heavy atoms may not be true. Indeed, the motion of heavy atoms on the surface is inhibited by the irregularities of the surface potential that result from numerous surface defects and the discrete position of the lattice molecules. The exact values for the corresponding potential minima are not known for the interstellar grains and we have to relay on the laboratory data for the materials which probably have many fewer defects than those in the ISM. Atoms trapped in lattice defects may enter a potential minimum with an energy comparable to that of chemisorption and lose their mobility for the conditions we discuss here. For physically adsorbed atoms to move between adjacent minima (a few Ångström) a barrier of $E_B$ height should be surmounted. $E_B$ is typically around one third of the total binding energy (although see Jaycock & Parfitt 1986) which is greater than 800 K for $O$, $N$ and $C$ atoms (Watson 1976). The probability of quantum tunnelling for heavy atoms is negligible and therefore their mobility is due to thermal hopping. The characteristic time for a particle to hop over a barrier is (Watson 1976)

$$t_h \approx \frac{1}{\nu_0} \exp\left(\frac{E_B}{kT_s}\right) \tag{33}$$

where $\nu_0 \approx 10^{12}$ s$^{-1}$ is a characteristic vibration frequency for the adsorbed particle. For $T_s = 7$ K, this gives $t_h \approx 10^5$ s. Since this is a random walk process, a time

$$t_M \approx \frac{1}{\nu_0} M^2 \exp\left(\frac{E_B}{kT_s}\right) \tag{34}$$

is required for a particle to get to a point $M$ lattice spacing away. This makes the diffusion time for a heavy atom inside an irregular fractal structure of a grain of $10^{-5}$ cm comparable with the times of grain-grain collisions, i.e. greater than $t_r$. Such situation is expected for a low abundance of $H$ atoms or when reaction between $H$ atoms and 'poisoners' are strongly inhibited. On the contrary, as long as $H$ atoms have a fractional abundance greater than about $10^{-3}$, there are enough $H$ atoms to hydrogenate, essentially instantaneously, every heavy atom that is accreted (see Millar 1993). The products of the reactions are expected either to be ejected as $OH$, $CH$, $CH_2$, $CH_3$, $NH$, $NH_2$ or remain on the grain to produce fully saturated forms like $H_2O$, $CH_4$, $NH_3$. These molecules are highly immobile and their diffusion inside pores is negligible. Therefore the probability of covering active sites even relatively close to the outer surface of a grain may be substantially reduced as compared to resurfacing of the outer parts of the grain. However, to dominate grain dynamics the the number of active sites within the grain up to the depth $l_{ac}$ should exceed the number of active sites over the outer surface.

Our discussion reveals that for fractal grains simple estimates for $t_L$ given in Spitzer & McGlynn (1979) are no more applicable. In fact, $t_L$ apart from the subtle details of grain surface chemistry may depend on grain fractal dimension as well as on $l$ and $l_{ac}$. To illustrate this phenomenon, we may assume

$$t_L = \varpi t_m \left(\frac{l}{r_0}\right)^{d-2} \left(\frac{2 \cdot 10^{-5} \text{ cm}}{l}\right) \tag{35}$$

where $\varpi$ is a parameter of the model, which is chosen 10 in accordance with the considerations above. Note, that our conjecture for $t_L$ is not applicable even qualitatively for small $l$. Indeed, if the time of grain collision with $H$ atom

$$\begin{aligned} t_c &= \frac{1}{\pi n_H v_H l^2} \\ &\approx 3 \cdot 10^2 \left(\frac{2 \cdot 10^{-5} \text{ cm}}{l}\right)^2 \left(\frac{20 \text{ cm}^{-3}}{n}\right) \left(\frac{T_g}{80 \text{ K}}\right)^{\frac{1}{2}} \text{ s} \end{aligned} \tag{36}$$

becomes less than the average time for an oxygen atom to scan the grain surface with $A_S$ sites (see

Tielens & Allamandola 1987):

$$t_s \approx \frac{A_S \nu_0^{-1}}{4\pi} \exp\left(\frac{E_B}{kT_S}\right)$$

$$\approx 3 \cdot 10^3 \left(\frac{l}{2 \cdot 10^{-5} \text{ cm}}\right)^2 \left(\frac{l}{r_0}\right)^{d-2}$$

$$\times \exp\left\{\left(\frac{E_B/k}{800 \text{ K}}\right)\left(\frac{10 \text{ K}}{T_S}\right)\right\} \text{ s} \qquad (37)$$

oxygen atoms poison sites of $H_2$ formation before a monomolecular layer is formed. For grain surfaces covered with a few monolayers of ice (see Zhang & Buch 1990, Buch & Zhang 1991) this may mean that oxygen occupies places which correspond to the potential minima, e.g. places of most probable $H_2$ formation. The critical size is

$$l_{cr} \approx 10^{-5} \left(\frac{l_{cr}}{r_0}\right)^{-\frac{d-2}{4}} \left(\frac{20 \text{ cm}^{-3}}{n}\right)^{\frac{1}{4}} \left(\frac{T_g}{80 \text{ K}}\right)^{\frac{1}{8}}$$

$$\times \exp\left\{-\frac{1}{4}\left(\frac{E_B/k}{800 \text{ K}}\right)\left(\frac{10 \text{ K}}{T_s}\right)\right\} \text{ cm} \qquad (38)$$

which gives $0.6 \cdot 10^{-5}$ cm for $l_{cr}/r_0 = 100$ and $d = 2.5$. Naturally, grains with a few active sites are much more $l_{cr}$ sensitive than those with high density of active sites. Indeed, if a grain with $l = 0.5 \cdot 10^{-5}$ cm has only $10^2$ active sites (Hollenbach & Salpeter 1971, Purcell 1979), it needs to collide just with a hundred $O$ atoms to cover them. This number becomes more than $3 \cdot 10^5$ if the grain is completely covered by the sites. For $l$ less than $l_{cr}$ we expect a drastic decrease in $t_L$, which is inversely proportional to the fraction of surface covered by active sites of $H_2$ formation and $t_L$ becomes much less than $t_m$. As soon as $t_L \ll t_m$, one has to use $t_L$ instead of $t_d$ in Eq. (6). Thus the 'regular' $\Omega$ falls much lower than 'thermal' $\omega_T$ [see Eq. (9)] consequently the rotation ceize to be suprathermal. The original Davis-Greenstein theory should be applied to obtain the measure of alignment [see Eqs (23) and (14)].

To summarise, it has been shown that $t_L$ depends on subtle processes of grain surface chemistry and for grains greater than $l_{cr}$, $t_L$ is expected to be greater than the estimate in Spitzer & McGlynn (1979) if grains are fractal.

### 4.3. 'Crossovers'

As the $z$ component of the torque acting on each suprathermally rotating grain (see section 3) changes its sign, the grain will be braked and then will start rotating around the axis which, in general, is not parallel to the initail axis of rotation. However, it was shown in Spitzer & McGlynn (1979) that a partial preservation of orientation during such crossovers is possible. This provides for the disorientation factor (see Appendix)

$$F \approx 0.8 \left(\frac{m_{H2} v_{H2}}{10^{-18} \text{ g cm s}^{-1}}\right)^{\frac{1}{3}} \left(\frac{\varrho}{3 \text{ g cm}^{-3}}\right)^{-\frac{1}{2}}$$

$$\times \left(\frac{l}{2 \cdot 10^{-5} \text{ cm}}\right)^{-\frac{4}{3}} \left(\frac{n_H}{20 \text{ cm}^{-3}}\right)^{-\frac{1}{6}} \left(\frac{v_H}{1.3 \cdot 10^5 \text{ cm s}^{-1}}\right)^{-1/6}$$

$$\times \left(\frac{\gamma_1}{0.2}\right)^{-\frac{1}{6}} \left(\frac{\alpha_{H2}}{4 \cdot 10^{10} \text{ cm}^{-2}}\right)^{\frac{1}{3}} \left(\frac{l_{ac}}{r_0}\right)^{\frac{d-2}{3}} \left(\frac{l_{ac}}{l}\right)^{-\frac{d}{6}} \qquad (39)$$

where $l_{ac}$ is assumed greater than $r_0$. The factor $(\alpha_{H2}/4 \cdot 10^{10} \text{ cm}^{-2})^{1/3} \pi l^2 (l_{ac}/r_0)^{(d-2)/3}$ is the number of active sites contributing to the spin-up. The disorientation grows with this number and decreases with the fraction of surface involved in 'torque generating', i.e. $\left(\frac{l_{ac}}{l}\right)^d$.

### 4.4. Grains completely covered by active sites

If a suprathermally rotating grain is completely covered by active sites and the uncompensated torque is due to the grain shape, this shape has to be substantially altered, e.g. due to grain-grain collisions, to change the direction of the torque. The corresponding time is, as a rule, greater than $t_r$.

At the same time, we have seen in section 3 that the degree of suprathermality is low for such grains. Therefore fluctuations due to the tail of the Maxwellian distribution are likely to influence them. This influence is expected to increase for large grains as they lose suprathermality due to $(l_{ac}/l)^d$ coefficient and for small grains due to a rapid growth of $\omega_T$ with the decrease of grain size. For some materials and fractal dimensions these sizes can overlap, and such grains cannot exhibit suprathermal behaviour. Their alignment should be described by the Davis-Greenstein theory [see Eqs (23) and (10)].

If the spatial variations of $H_2$ formation are due to physico-chemical properties of grain surface, crossovers are expected as a result of resurfacing and our earlier estimates for $t_L$ are applicable, provided that grains are suprathermal.

### 4.5. Grain with a low density of active sites

If there are only a few active sites over grain surface, the accretion of a monomolecular layer alters substantially grain rotation. For $F < 1$ and $l > l_{cr}$, one obtains

$$\delta_{eff} = \frac{t_x}{Ft_r} \approx 0.03 \left(\frac{\varrho}{3 \text{ g cm}^{-3}}\right)^{1/2} \left(\frac{l}{2 \times 10^{-5} \text{ cm}}\right)^{1/3}$$

$$\times \left(\frac{n}{20 \text{ cm}^{-3}}\right)^{-1} \left(\frac{n_H}{20 \text{ cm}^{-3}}\right)^{1/6} K^{-1}(\Omega) \left(\frac{\gamma_1}{0.2}\right)^{1/6}$$

$$\times \left(\frac{v_a}{1.3 \cdot 10^5 \text{ cm s}^{-1}}\right)^{-1} \left(\frac{B}{2 \cdot 10^{-6} \text{ G}}\right)^2$$

$$\times \left(\frac{15 \text{ K}}{T_S}\right) \left(\frac{v_H}{1.3 \cdot 10^5 \text{ cm s}^{-1}}\right)^{1/6} \left(\frac{l_{ac}}{r_0}\right)^{-\frac{d-2}{3}} \left(\frac{l_{ac}}{l}\right)^{\frac{d}{6}}$$

$$\times \left(\frac{\alpha_{H2}}{4 \cdot 10^{10} \text{ cm}^{-2}}\right)^{-1/3} \left(0.6 + \frac{t_L}{t_m}\right) \qquad (40)$$

If $t_L$ is given by Eq. (35), it is possible to plot the alingment measure as a function of $l$ (see Fig. 3).

It is worth noting that the fractal dimension can vary with grain composition, size and physical conditions around. For instance, grain pores can be covered by ice as a result of mantle growth. If $t_L \gg t_m$, the alignment measure for large grains decreases less rapidly because of partial preservation of the angular momentum direction during crossovers which results in $F < 1$ (see Fig. 3). The increase of the said measure of claimed in Spitzer & McGlynn (1979) is possible for $t_L \leq t_m$. The discontinuity of the derivative that is obvious from Fig. 3 is an artifact of the simplified model adopted. Indeed, even for $F > 1$ there should be a residual preservation of alignment, while in our model a complete disorientation after the threshold $F = 1$ is assumed. In fact, one should bear in mind that Fig. 3 is obtained within a model which adopts a conjecture for $t_L$ given by Eq. (35). Further research should be directed to obtaining a more realistic analytical approximation for the time of spin-up $t_L$.

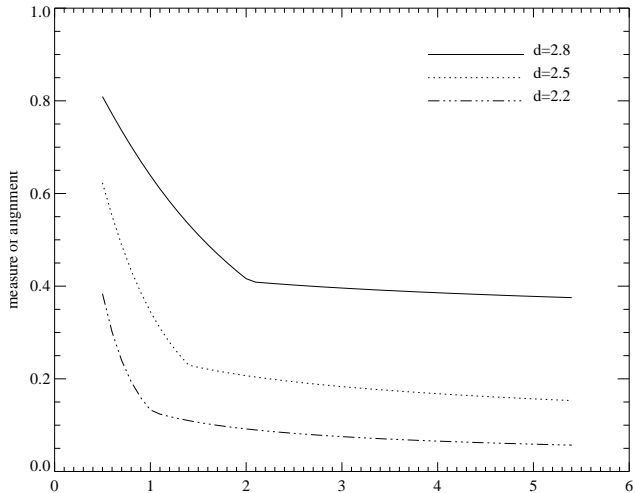

**Fig. 3.** The alignment measure as a function of grain size $l$ in units of $2 \cdot 10^{-5}$ cm for $r_0 = 2 \cdot 10^{-6}$ cm and different fractal dimensions $d$. We intentionally start from $l = 10^{-5}$ cm, as for smaller grains we expect the decrease of the alignment measure, which depends on grain physics and chemistry.

## 5. Theory & observations

### 5.1. Small and large grains

Using the MNR grain model (Mathis et al.1977), it is possible to show (Mathis 1979) that a good fit of the polarization curve may be obtained if grains with radii above a certain threshold are aligned [although see Johnson (1982), where the observational evidence for better alignment of small grains is claimed]. The above dependence was accounted for in Mathis (1986) by assuming that only those grains that contain at least one superferromagnetic cluster are being aligned. Within the adopted model, the probability of having such a cluster is greater for larger grains. Alternatively, for the Gold-type alignment (Lazarian 1994a), it is possible to show that only grains larger than $10^{-5}$ cm are sufficiently inertial to be aligned under Alfvénic perturbations in the typical ISM conditions.[4] Here we are interested whether alignment of suprathermally rotating fractal grains can reproduce this type of behaviour for the polarization curve, and thus need to discuss the alignment of large and small grains separately.

---
[4]Grains are expected to be aligned with their long axis perpendicular to magnetic field lines, which corresponds to observations (Lazarian 1994a).

Our calculations in section 3 have shown that if the density of active sites over grain surface is high, such grains are likely to lose the suprathermal advantage if their sizes are less than $2 \cdot 10^{-6}$ cm. At the same time, our calculations in section 4 indicate that if there are only a few sites of $H_2$ formation per grain, poisoning mainly by $O$ atoms intensifies for $l$ less than $l_{cr} \approx 6 \cdot 10^{-6}$ cm. Therefore whatever the density of active sites, grains less than a particular size are likely to be aligned only marginally.

The critical size $l_{cr}$ can increase due to the inverse dependence of $T_S$ on $l$ (Greenberg 1971):

$$T_S(l) \approx T_S \left( \frac{2 \cdot 10^{-5} \text{ cm}}{l} \right)^{0.2} \text{ K} \qquad (41)$$

where $T_S$ is the temperature of a grain of size $2 \cdot 10^{-5}$ cm. This is a result of suppressed thermal emission from small grains in infrared. Although the power law index is just 0.2, this dependence may be essential in the view of exponential dependence of $l_{cr}$ on temperature [see Eq. (38)].

Another important factor is the difference between the imaginary parts of magnetic susceptibilities of slowly and rapidly rotating grains. Since the relaxation time for $\Omega > 10^9$ s$^{-1}$ is proportional to $\Omega$ (see section 3), $\delta_1 \sim (l\Omega)^{-1}$. As $\Omega$ increases as $l^{-(2+d)}$ with the decrease of $l$, $\delta_1$ should decrease at least as $l^{-1}$. Thus, we expect to see a decrease of alignment for $l$ smaller than a critical value. For a grain with a few hundreds of active sites, this corresponds to $l < 10^{-5}$ cm.

Note, that even for thermally driven rotation given by Eq. (9) we expect the imaginary part of magnetic susceptibility to come to the saturation for $l \sim 10^{-6}$ cm. Therefore $\delta_1$ given by Eq. (21) decreases for smaller $l$ as $l^{3/2}$. In fact, suprathermal rotation for some range of grain sizes and some materials can suppress alignment instead of increasing it. As the rate of accretion is proportional to $l^{-1}$ [see Eq. (30)], it is easy to show using Eqs (8) and (18) that $\delta_{eff}$ given by Eq. (26) decreases as $l^{(d-2)/2}$ for $\Omega > \omega_{cr}$. The combination of this effect with enhanced poisoning for small grains makes the decrease of the alignment measure with $l$ much more rapid. The variations in concentration of paramagnetic ions, otherwise marginally important,[5] may result in substantial changes in the sizes of aligned suprathermally rotating grains. Similarly, the decrease of atomic hydrogen concentration which mitigates the degree of suprathermality can in some cases increase alignment for a particular range of grain sizes.

Other factors that can contribute to a lower degree of polarization from small grains may be their shape and the decreased efficiency of $H_2$ formation over their surfaces. As for the shape, it is impossible to exclude that small grains become more spherical and the alignment of their angular momenta does not affect the overall polarization. As for the formation of $H_2$, it is inefficient on PAH surfaces (see Tielens 1993). This property may be shared by some carbonaceous grains if, according to Duley & Williams (1988), they consist of loosely connected PAH structures.

---
[5]It is stated in Jones & Spitzer (1967) that in the limit of low frequency, the imaginary part of the magnetic sucseptibility is independent of the magnitude of individual magnetic moments. Therefore in this limit, nuclear paramagnetism should be as effective as electronic paramagnetism in relaxing the interstellar grains.

To summarise, for suprathermally rotating grains the degree of alignment is expected to be low for small grains.

A quantitative comparison of theory and observations for polarisation produced by dust with a power-law size distribution and with the grain elongation varying with size will be given elsewhere.

### 5.2. Silicate and carbonaceous grains

Excess in polarization associated with 9.7 $\mu$m silicate feature demonstrates that silicate grains are aligned in both low and high density regions of the ISM (see Whittet 1992). The general form of the polarization curve with its smooth peak in visible and steady decline in infrared is difficult to reproduce using metals or other strong absorbers such as graphite. However, it seems that carbonaceous grains discussed in Duley (1993) can be aligned. These grains have a refractive index $\bar{m} = n + ik \geq 1$ and have a porous non-absorbing matrix with a dilute dispersion of absorbing molecules. It was shown in Duley (1993) that a polyethylene matrix should not contain more than $4 \cdot 10^{19}$ cm$^{-3}$ molecules of coronene $C_{24}H_{24}$ for $k$ be less than 0.1. Such a solid satisfies the requirement $\bar{m} \simeq 1 - 1.7$ and allows efficient utilisation of carbon to produce extinction. At the same time, fluffy carbonaceous structures which may be partially graphitized by exposure to the interstellar radiation field are unlikely to be aligned by the Purcell's mechanism (1975, 1979) as the ejected H$_2$ molecules are expected to have little kinetic energy. We argue in our next paper (Lazarian 1994c that due to peculiarities of H$_2$ formation over grain surface (Duley & Williams 1986, 1993) grains with surfaces of H$_2$O ice, defected silicate or polymeric carbonaceous material are likely to exibit an enhanced alignment due to suprathermal rotation, while those of aromatic carbonaceous material or graphite are not.

### 5.3. Cold and warm grains

We have already seen that even small changes of temperature can substantially alter grain alignment. In fact, an increase of temperature over a few degrees considerably increases $l_{cr}$ due its exponential dependence on temperature. As a result, poisoning atoms become sufficiently mobile to attack active sites directly and therefore the time of changing position of active sites becomes less than the time of accreting a monomolecular layer.

If the temperature is further increased, the $H_2$ formation becomes suppressed. For a reaction between physically and chemically adsorbed $H$ atoms to occur, an activation barrier should be overcome. A critical temperature $T_{cr}$ can be defined for which the probability of reaction with a chemisorbed atom is equal to the rate of evaporation:

$$T_{cr} = \frac{E_b}{k \ln(\tau_r \nu_0)} \qquad (42)$$

where $\tau_r$ is the reaction time-scale $\sim 10^{-6}$ s. Estimates provide $T_{cr} \approx 50$ K for $H_2$ formation over a silicate surface (Tielens & Allamandola 1987). This means that, for instance, the aligned warm dust in the core regions of OMC-1 (see Chrysostomou et al. 1994) can not rotate suprathermally due to $H_2$ formation. Therefore it is unlikely that the suprathermal paramagnetic alignment of warm ($T_S > 50$ K) dust be efficient. Thus it either the Gold-type alignment (see Lazarian 1994a) or Davis-Greenstein mechanism that drives the alignment of warm dust.

### 5.4. Photodesorption

Throughout our discussion we assumed high values of grain resurfacing and therefore high rates of changing locations of active sites. This may not always be the case with grains in diffuse clouds. For instance, some observations (see Tanaka et al. 1990) testify that ice mantles occur only towards sources with visual extinction $A_v > 10$ (although see Whittet et al. 1988). This may correspond to cleaning grain surfaces by photodesorption, and therefore the existence time for Purcell 'rockets' may be greater than that given by Eq. (30). Indeed, the strongest poisoners for active sites are oxygen, nitrogen and carbon as they are usually adsorbed into monomolecular layer in time much shorter than $t_r$. However, $H$ atoms are expected to tunnel beyond the activation barrier for energetically favourable reactions of $H_2O$, $NH_3$ and $CH_4$ formation. These molecules are likely to be cleaned from grain surfaces by photodesorption which has a characteristic time $\sim 10^9$ s for an unshielded regions of diffuse clouds. This time is usually less than $t_r$, i.e. a better alignment may be achieved in comparison with the predictions we have obtained in the present paper.

However, we do not have conclusive observational evidence in favour of the picture discussed above. On the contrary, elaborate modelling of dust alignment on the line of sight to the Becklin-Neugebauer (BN) object in Lee & Draine (1985) seems to testify that the core mantle grains tend to be well aligned. An evidence that the degree of polarization is correlated with the strength of 3.1 $\mu$m water ice band follows from observations (see Joyce & Simon 1982) although this is by no means a universal tendency (Dyck & Lonsdale 1981). This seems to contradict the notion that keeping surfaces of grains clean considerably improves their alignment (compare Johnson 1982). At the same time, we probably do not observe significant depletion of elements heavier than $He$ in clouds (see Whittet 1992) to attribute the decrease of poisoning to the fact that all the 'troublesome poisoners' are adsorbed by grains. Therefore the possibility of the prolonged spin-up due to the fractal structure of grains may be valuable for explaining observations.

At the same time, indications that alignment is systematically less efficient towards stars with higher extinction (Tamura et al. 1987, Whittet et al. 1994) as well as low degree of polarization in far infrared reported in Goodman (1994) may testify that the alignment decreases as major part of atomic hydrogen is converted into molecular form.

## 6. Conclusions

The main conclusion of this study is that the fractal structure of grains influences the efficiency of paramagnetic alignment. The rotation of grains with a developed fractal surface becomes less vulnerable to the resurfacing induced by accretion and this improves the alignment. At the same time, fractality can inhibit alignment if the surface coverage by active sites is high and the majority of $H_2$ molecules thermolise within fractal pores before leaving grains. Our study reveals a non-monotonous dependence of the alignment measure on grain size. The details of the said dependence are influenced by binding energy, the number of active sites, grain composition and their temperature. For a low density distribution of active sites there exist a critical size caused by active poisoning and, for grains less than this size, the alignment is inhibited due to frequent crossovers. Yet another 'low' critical size is shown to exist due to a limited value of the imaginary part of magnetic susceptibility. This weakens the alignment of rapidly rotating small grains. Considering a high density of active sites, we found that grains both less and greater than particular sizes loose their suprathermality, and the original Davis-Greenstein theory should be applied to describe their alignment. In both cases, the alignment is temperature dependent and for warm grains ($T_S > 50$ K) the 'chemically driven' suprathermal rotation is unlikely. A further study of alignment for different ISM regions should give an insight into complex processes of grain physics and chemistry.

**Acknowledgement**

This paper owes much to my discussions with P. Myers, M. Rees, J. Rollings, L. Spitzer and D. Williams. I take a great pleasure to thank J. Mathis for numerous helpful remarks and suggestions which considerably improved the quality of this paper. Encouragement of N. Weiss is gratefully acknowledged. I take a great pleasure to thank the Institute of Astronomy, University of Cambridge, for the Isaac Newton Scholarship they generously awarded me, which provided financial support for this work.

**References**

Aannestad P.A. & Greenberg J.M., 1983, *ApJ*, **272**, 551.
Aron J. & Max C., 1975, *ApJ*, **196**, L77.
Buch V. & Zhang Q., 1991, *ApJ*, **379**, 647.
Cardelli J.A., Savage B.D., Bruhweiler F.C., Smith A.M., Ebbets D., Sembach K.K. & Sofia U.J., 1991, *ApJ*, **377**, L57.
Chrysostomou A., Hough J.H., Burton M.G. & Tamura M., 1994, *MNRAS*, **268**, 325.
Clayton G.C. et al., 1992, *ApJ*, **382**, L83.
Codina-Landaberry S., Marcondes-Machado J.A. & Magalhaes A.M., 1974, *A&A*, **36**, 173.
Davis J. & Greenstein J.L., 1951, *ApJ*, **114**, 206.
Dolginov A.Z. & Mytrophanov I.G., 1976, *Ap&SS*, **43**, 291.
Duley W.W., 1978, *ApJ*, **219**, L129.
Duley W.W., 1993, in Dust and Chemistry in Astronomy, Millar T.J. & Williams D.A. (eds.), Bristol Inst. Physics, 143.
Duley W.W. & Williams D.A., 1986, *MNRAS*, **223**, 177.
Duley W.W. & Williams D.A., 1988, *MNRAS*, **231**, 969.
Duley W.W. & Williams D.A., 1993, *MNRAS*, **260**, 37.
Dyck H.M. & Lonsdale C.J., 1981, in IAU Simposium-96, Infrared Astronomy, C.G. Wynn-Williams & D.P. Cruikshank (eds.), Dordercht: Reidel, 223.
Elmegreen B.G., 1992, in *The Galactic Interstellar Medium*, D. Pfenniger & P. Bartholdi (eds.), Springer-Verlag, p. 157.
Gold T., 1951, *Nat*, **169**, 322.
Gold T., 1952, *MNRAS*, **112**, 215.
Goodman A.A., 1994, in The Physicss and Chemistry of Interstellar Molecular Clouds, Proc. of the 2nd Cologne-Zermatt Symposium (in press)
Gradstein I.S. & Ryzhik I.M., 1965, Table of Integrals, Series and Products, Academic Press, New York.
Greenber J.M., 1971, *A&A*, **12**, 240.
Hiltner W.A., 1949, *ApJ*, **109**, 471.
Hollenbach D. & Salpeter E.E., 1970, *J.Chem.Phys.*, **53**, 79.
Hollenbach D. & Salpeter E.E., 1971, *ApJ*, **163**, 155.
Jaycock M.J & Parfitt G.D., 1986, in Chmistry of Interfaces, Wiley and Sons, new York.
Johnson P.E., 1982, *Nature*, **295**, 371.
Jones R.V. & Spitzer L.,Jr, 1967, *ApJ*, **147**, 943.
Jones A.P. & Williams D.A., 1984, *MNRAS*, **209**, 955.
Joyce R.R. & Simon T. 1982, *ApJ*, **260**, 604.
Lazarian A., 1992, *Astron. and Astrophys. Transactions*, **3**, 33.
Lazarian A., 1993, in Proc. of the 8th Manchester Meeting '*Kinematics and Dynamics of Diffuse Astrophysical Media*', Manchester, 22-26 March.
Lazarian A., 1994a, *MNRAS*, **268**, 713.
Lazarian A., 1994b, *MNRAS* (submitted paper)
Lazarian A., 1994c, *MNRAS* (submitted paper)
Lee H.M. & Draine B.T., 1985, *ApJ*, **290**, 211.
Lucas D. & Ewing G.E., 1981, *Chem. Phys.*, **58**, 385.
Lume K. & Rahola J., 1994, *ApJ*, **425**, 653.
Mandelbrot B.B., 1983, The Fractal Geometry of Nature, Freeman, 1982.
Martin P.G., 1978, *Cosmic Dust. Its Impact on Astronomy*, Clarendon Press, Oxford.
Mathis J.S., 1979, *ApJ*, **232**, 747.
Mathis J.S., 1986, *ApJ*, **308**, 281.
Mathis J.S., 1990, *Ann. Rev. Astron. Astrophys.*, **28**, 37.
Mathis J.S., Rumpl W. & Nordsieck K.H., 1977, *ApJ*, **217**, 425.
Mathis J.S. & Whiffen G., 1989, *ApJ*, **341**, 808.
Millar T.J., 1993, in Dust and Chemistry in Astronomy, Millar T.J. & Williams D.A. (eds.), Bristol Inst. Physics, p. 249.
Myers P.C., 1987, in *Interstellar Processes*, eds D.J. Hollenbach and H.A. Thronson, Jr, Kluwer, Dordrecht, Reidel, p 71.
Myers P.C., 1994, in 'The Structure and Content of Molecular Clouds', eds Wilson T.L. & Johnston K.J., Berlin, Springer-Verlag.
Purcell E.M., 1975, in *The Dusty Universe*, eds G.B. Field & A.G.W. Cameron, New York, Neal Watson, p. 155.
Purcell E.M., 1979, *ApJ*, **231**, 404.
Purcell E.M. & Spitzer L., Jr, 1971, *ApJ*, **167**, 31.
Roberge W.G., DeGraff T.A. & Flaherty J.E., 1993, *ApJ*, **418**, 287.
Sorrell W.H., 1994, *MNRAS*, **268**, 40.
Spitzer L., Jr, 1978, *Physical Processes in the Interstellar Medium*, Wiley-Interscience Publ., New York.
Spitzer L., Jr & McGlynn T.A., 1979, *ApJ*, **231**, 417.
Tamura M., Nagata T., Sato S. & Tanaka M., 1987, *MNRAS*, **224**, 413.
Tanaka M., Sato S., Nagata T. & Yamamoto T., 1990, *ApJ*, **352**, 724.
Tielens A.G.G.M., 1993, in Dust and Chemistry in Astronomy, Millar T.J. & Williams D.A. (eds.), Bristol Inst. Physics, p. 103.
Tielens A.G.G.M., Allamandola L.J., 1987, in Interstellar Processes, eds. D.J. Hollenbach and Thronson, Reidel, Dordrecht, p. 397.
Wagenblast R., Williams D.A., Millar T.J. & Nejad L.A.M., 1993, *MNRAS*, **260**, 420.
Watson W.D., 1976, in Atomic and Molecular Physics and the Interstellar Matter, R. Balian, P. Encrenaz & J. Lequeux (eds.), Elsevier, Asmterdam, 177.
Whittet D.C.B., 1992, *Dust in the Galactic Environment*, Bristol Inst. Physics.

Whittet D.C.B., Bode M.F., Longmore A.J., Adamson A.J., McFadzean A.D., Aitken D.A. & Roche P.F., 1988, *MNRAS*, **233**, 321.
Whittet D.C.B., Gerakines P.A., Carkner A.L., Hough J.H., Martin P.G., Prusti T. & Kilkenny D., 1994, *MNRAS*, **268**, 1.
Wilking B.A., Lebofsky M.J. & Rieke G.H., 1982, *A.J.*, **87**, 695.
Williams D.A., 1988, in Dust in the Universe, Bailey M.E. & Williams D.A. (eds), CUP, p. 391.
Williams D.A., 1993, in Dust and Chemistry in Astronomy, Millar T.J. & Williams D.A. (eds.), Bristol Inst. Physics, p. 71.
Williams D.A., Hartquist T.W. & Whittet D.C.B., 1992, *MNRAS*, **258**, 599.
Witten T.A. & Cates M.E., 1986, *Science*, **232**, 1607.
Wright E.L., 1987, *ApJ*, **320**, 818.
Zhang, Q. & Buch V., 1990, *J. Chem. Phys.*, **92**, 5004.

## A. Disorientation parameter for fractal grains

A partial preservation of alignment during crossovers can be described using parameter $F$ introduced in Spitzer & McGlynn (1979) as

$$\cos\langle\chi_T\rangle = \exp(-F) \tag{A1}$$

where $\chi_T$ is the angle of deviation of the angular momentum. $F$ can also be expressed as an integral over the mean squared deviations due to elementary torques

$$F \equiv \frac{1}{2}\int_{-\infty}^{+\infty} N\langle(\triangle\chi_i)^2\rangle \mathrm{d}t \tag{A2}$$

where $N$ is the frequency of the torque events. Calculations in Spitzer & McGlynn (1979) give

$$F = \frac{\pi}{4}\left\{\frac{\langle(\triangle J_z)^2\rangle}{|J_\perp|\langle\triangle J_z\rangle}\left(1 + \frac{3\langle(\triangle J_\perp)^2\rangle}{2\langle(\triangle J_z)^2\rangle}\right)\right\} \tag{A3}$$

in the approximation of the constant angular momentum component that is perpendicular to the axis of the greatest inertia. Note, that $\triangle J_z$ and $\triangle J_\perp$ are elementary changes of the angular momentum in an individual torque event, and the angular brackets denote averaging over distributions of $\triangle J_z$ and $J_\perp$.

The squared dispersion of $J_\perp$ at $t=0$ is given in Spitzer & McGlynn (1979, eq. 42):

$$\langle J_\perp^2\rangle(0) \approx \left(\frac{3}{2}\right)^{1/3}\Gamma\left(\frac{4}{3}\right)(A_a N)^{1/3}\langle\triangle J_z\rangle^{-2/3}I_z^{2/3}$$
$$\times \langle(\triangle J_\perp)^2\rangle \tag{A4}$$

where $\Gamma(x)$ is the Gamma function, and the coefficient of the Barnett relaxation (Purcell 1979, eq. 46)

$$A_a = 6\times 10^{17}\left(\frac{l}{2\times 10^{-5}\ \mathrm{cm}}\right)^2\left(\frac{\varrho}{3\ \mathrm{g\ cm^{-3}}}\right)\frac{T_S}{15\ K}\ \mathrm{s}^{-1} \tag{A5}$$

is assumed.

The frequency of the torque events is proportional to the frequency of grain collisions with $H$ atoms multiplied by $(l_{ac}/l)^d$, i.e.

$$N \approx 5\times 10^{-4}\left(\frac{\gamma_1}{0.2}\right)\left(\frac{l}{10^{-5}\ \mathrm{cm}}\right)\left(\frac{n_H}{20\ \mathrm{cm^{-3}}}\right)$$
$$\times \left(\frac{l_{ac}}{l}\right)^d\left(\frac{v_H}{10^5\ \mathrm{cm\ s^{-1}}}\right)\ \mathrm{s}^{-1} \tag{A6}$$

To calculate $F$ from Eq. (A3), one needs to know the average of $J_\perp^{-1}(0)$. Note, that $\langle J_\perp^2(0)\rangle$ is a sum of $\langle J_x^2(0)\rangle$ and $\langle J_y^2(0)\rangle$ and due to the symmetry inherent to the problem

$$\alpha_{H2}^2 = \langle J_x^2(0)\rangle = \langle J_y^2(0)\rangle = 0.5\langle J_\perp^2(0)\rangle \tag{A7}$$

For a Gaussian distribution, one gets

$$\frac{1}{\langle J_\perp^2(0)\rangle} = \frac{1}{2\pi\alpha_{H2}}\iint_{-\infty}^{+\infty}\frac{1}{\sqrt{x^2+y^2}}$$
$$\times \exp\left\{-\frac{x^2+y^2}{2\alpha_{H2}^2}\right\}\mathrm{d}x\mathrm{d}y \tag{A8}$$

which can be calculated in polar coordinates to give

$$\frac{1}{\langle J_\perp^2(0)\rangle} = \sqrt{\frac{\pi}{2}}\frac{1}{\alpha_{H2}} = \frac{\sqrt{\pi}}{\langle J_\perp^2(0)\rangle^{1/2}} \tag{A9}$$

Similarly the Gaussian averages can be found for $[\triangle J_z]^{-2/3}$, where the square brackets $[...]$ denote averaging over the surface of an individual grain. The dispersion of $[\triangle J_z]$ which we will denote $\sigma_2$ is inversely proportional to the square root of the number of active sites $\nu_{ac}$ over the grain surface and depends on grain geometry and distribution of active sites in respect to the $z$-axis. Therefore

$$\langle[\triangle J_z]^{-2/3}\rangle = \frac{1}{\sigma_2\sqrt{2\pi}}\int_{-\infty}^{+\infty}\frac{1}{x^{2/3}}\exp\left\{-\frac{1}{2}\frac{x^2}{\sigma_2^2}\right\}\mathrm{d}x \tag{A10}$$

which gives (see Gradshtein & Ryzhik 1965, [3.462(1)] and [9.241(1)])

$$\langle[\triangle J_z]^{-2/3}\rangle = \frac{1}{\sigma_2^{\frac{2}{3}}}\left\{\sqrt{\frac{2}{\pi}}\Gamma\left(\frac{1}{3}\right)D_{-1/3}(0)\right\} \tag{A11}$$

where $D_p(x)$ is the parabolic cylinder function and $\Gamma(x)$ is the Gamma function.

Thus

$$F = \tilde{\kappa}\frac{\langle(\triangle J_z)^2\rangle}{(A_a N)^{1/6}I_z^{1/3}\langle(\triangle J_z)^2\rangle^{1/3}\langle(\triangle J_\perp)^2\rangle^{1/2}}$$
$$\times \left(1 + \frac{3\langle(\triangle J_\perp)^2\rangle}{2\langle(\triangle J_z)^2\rangle}\right) \tag{A12}$$

where the numerical factor is equal to

$$\tilde{\kappa} = \left(\frac{3}{4}\right)^{-1/6}\Gamma^{-1/2}(4/3)\int_0^\infty e^{-\frac{x^2}{2}}x^{-2/3}\mathrm{d}x \tag{A13}$$

Assuming $\langle(\triangle J_z)^2\rangle = \langle(\triangle J_\perp)^2\rangle$, one obtains

$$F = \frac{5}{2}\tilde{\kappa}\frac{\langle(\triangle J_z)^2\rangle^{1/2}}{(A_a N)^{1/6}I_z^{1/3}\langle(\triangle J_z)^2\rangle^{1/3}} \tag{A14}$$

Substituting $\langle(\triangle J_z)^2\rangle = \frac{l^2}{9} m_{H2}^2 v_{H2}^2$ as well as expressions for $A_a$, $N$ and $\langle(\triangle J_z)^2\rangle$ into Eq. (A14), one gets

$$F \approx 0.8 \left(\frac{m_{H2} v_{H2}}{10^{-18} \text{ g cm s}^{-1}}\right)^{1/3} \left(\frac{\nu_{ac}}{50}\right)^{1/3}$$

$$\times \left(\frac{\varrho}{3 \text{ g cm}^{-3}}\right)^{-1/2} \left(\frac{l}{2\times 10^{-5} \text{ cm}}\right)^{-2} \left(\frac{n_H}{20 \text{ cm}^{-3}}\right)^{-1/6}$$

$$\times \left(\frac{v_H}{1.3\times 10^5 \text{ cm s}^{-1}}\right)^{-1/6} \left(\frac{\gamma_1}{0.2}\right)^{-1/6} \left(\frac{l_{ac}}{l}\right)^{-d/6} \quad (A15)$$

The number of active sites $\nu_{ac}$ over the grain surface up to the depth $l_{ac}$ may be assumed to be proportional to the physical surface of the grain $S_g(l)$. Therefore for a fractal grain one obtains

$$F \approx 0.8 \left(\frac{\varrho}{3 \text{ gcm}^{-3}}\right)^{-1/2} \left(\frac{l}{2\times 10^{-5} \text{ cm}}\right)^{-4/3}$$

$$\times \left(\frac{n_H}{20 \text{ cm}^{-3}}\right)^{-1/6} \left(\frac{\gamma_1}{0.2}\right)^{-1/6} \left(\frac{l_{ac}}{l}\right)^{-d/6} \left(\frac{l_{ac}}{r_0}\right)^{\frac{d-2}{3}}$$

$$\times \left(\frac{v_H}{1.3\times 10^5 \text{ cm s}^{-1}}\right)^{-1/6} \left(\frac{\alpha_{H2}}{4\times 10^{10} \text{ cm}^{-2}}\right)^{1/3} \quad (A16)$$

which shows that $F$ depends on the grain fractal dimension. If $l_{ac}$ is much greater than $r_0$ and $d > 2$, $F$ increases which means that a partial preservation of alignment during crossover becomes marginal for small grains. However, as the number of active sites over the depth $l_{ac}$ decreases, the preservation of alignment becomes more important.